\documentstyle[12pt]{article}

\newlength{\extraspace}
\newlength{\extraspaces}
\newcommand{\be}{\begin{equation}}
\newcommand{\ee}{\end{equation}}
\newcommand{\bear}{\begin{eqnarray}}
\newcommand{\eear}{\end{eqnarray}}
\newcommand{\ba}{\begin{array}}
\newcommand{\ea}{\end{array}}
\newcommand{\vev}{\mbox{$\langle \rho \rangle$}}
\newcommand{\trace}{\mbox{\rm Tr}}

\newcommand{\pauli}{\mbox{$\vec{\tau}$}}

\newcommand{\vw}{\mbox{$\vec{w}$}}

\newcommand{\pdmu}{\mbox{$\partial_{\mu}$}}
\newcommand{\PDmu}{\mbox{$\partial^{\mu}$}}
\newcommand{\su}{\mbox{$SU(2)\times U(1)$}}

\newcommand{\lr}{\mbox{$SU(2)_{L}\times SU(2)_{R}$}}

\newcommand{\gev}{\mbox{\rm GeV}}
\newcommand{\wz}{\mbox{$w^{+} w^{-} \rightarrow z z$}}

\newcommand{\yu}{\mbox{$\Sigma$}}
\newcommand{\xee}{\mbox{$\xi$}}
\newcommand{\xpr}{\mbox{$\xi^{\prime}$}}
\newcommand{\xdpr}{\mbox{$\xi^{\prime\prime}$}}
\newcommand{\ctri}{\mbox{$\lambda_{3}$}}
\newcommand{\ctes}{\mbox{$\lambda_{4}$}}
\newcommand{\cA}{{\cal A}} 
 \newcommand{\cO}{{\cal O}}

\newcommand{\cL}{{\cal L}} 
\newcommand{\cM}{{\cal M}} 
\newcommand{\np}{Nucl.\ Phys.\ {\bf B}}
\newcommand{\pr}{Phys.\ Rev.\ }
\newcommand{\prd}{Phys.\ Rev.\ {\bf D}}
\newcommand{\prp}{Phys.\ Rep.\ }
\newcommand{\prl}{Phys.\ Rev.\ Lett.\ }
\newcommand{\pl}{Phys.\ Lett.\ {\bf B}}
\newcommand{\jsp}{J.\ Stat.\ Phys.\ }
\newcommand{\cmp}{Comm.\ Math.\ Phys.\ }

\begin{document}
\pagestyle{empty}

\begin{titlepage}
\begin{flushright}
{\rm BUHEP-93-7 \\ hep-ph/9304293 \\ \today}
\end{flushright}
\vspace{24pt}
\begin{center}
{\LARGE Phenomenology of a Non-Standard Higgs}\\
\vspace{40pt}
{\rm R. Sekhar Chivukula}\footnote{e-mail address: sekhar@weyl.bu.edu},
{\rm Vassilis Koulovassilopoulos}\footnote{e-mail address: vk@budoe.bu.edu}
\vspace*{0.5cm}

{\it Physics Department, Boston University,\\
590 Commonwealth Avenue,\\
Boston, MA 02215 USA}
\vskip 2.1cm
\rm
\vspace{25pt}
{\bf ABSTRACT}
\vspace{12pt}
\end{center}
\baselineskip=18pt
\begin{minipage}{15.5cm}

  The one-Higgs-doublet standard model is necessarily incomplete because of
  the triviality of the scalar symmetry-breaking sector.  If the Higgs mass is
  approximately 600 GeV or higher, there must be additional dynamics at a
  scale $\Lambda$ which is less than a few TeV. In this case the properties of
  the Higgs resonance can differ substantially from those predicted by the
  standard model.  In this letter we construct an effective Lagrangian
  description of a theory with a non-standard Higgs boson and analyze the
  features of a theory with such a resonance coupled to the Goldstone Bosons
  of the breaking of $\su$.  This Lagrangian describes the most general theory
  in which the Higgs and the Goldstone Bosons are the only particles with a
  mass small compared to $\Lambda$.  We compute the leading corrections to the
  decay width of the Higgs boson and the contribution to the Peskin-Takeuchi
  parameter $S$ and present results for the corrections to Goldstone Boson
  scattering. We find that the most prominent effects are due to one new
  parameter, which is directly related to the Higgs boson width.

\end{minipage}

\vfill
\end{titlepage}
\pagebreak
\baselineskip=18pt
\pagestyle{plain}
\setcounter{page}{1}

The standard one-doublet Higgs model predicts the existence of a neutral
scalar particle, the Higgs boson, whose couplings are related to its mass or
to the masses of the electroweak gauge bosons so as to preserve
renormalizability.  However, there are strong indications that, at least in
the limit of vanishing gauge and Yukawa couplings, the scalar
symmetry-breaking sector is trivial \cite{kn:phi4,kn:o(n)}.  This means that
in the continuum limit and for acceptable values of the bare scalar
self-coupling $\lambda_0$, the renormalized self-coupling $\lambda_R$ is {\it
  zero}. This implies that the standard one-doublet Higgs model of electroweak
symmetry breaking is necessarily incomplete and can only be viewed as a
low-energy effective theory below some finite cutoff $\Lambda$.

If the Higgs mass is approximately 600 GeV or higher, there must be additional
dynamics at a scale which is less than a few TeV \cite{kn:dn}: {\it i.e.} the
cutoff is low. In this case the properties of the Higgs resonance can differ
substantially from those predicted by the standard model.  In this letter we
construct an effective Lagrangian description of a theory with a non-standard
Higgs boson, and explore the properties of a theory with such a resonance
coupled to the Goldstone Bosons of the breaking of $\su$.  This Lagrangian
describes the most general (custodially symmetric \cite{kn:csu2}) theory in
which the Higgs and the Goldstone Bosons are the only particles with masses
small compared to $\Lambda$. We compute the leading corrections to the
decay width of the Higgs boson and to the Peskin-Takeuchi parameter $S$
\cite{kn:peskin}. We also present results for the corrections to Goldstone
Boson scattering. Details of the calculations and an explicit model with a
non-standard Higgs resonance will be presented elsewhere \cite{kn:vklong}.

The language of effective Lagrangians has been used extensively
\cite{kn:dobado} to study a strongly-interacting symmetry-breaking sector. One
usually assumes that the longitudinal gauge-bosons are the {\it only}
particles in the symmetry-breaking sector with masses of less than a TeV and,
therefore, (assuming a custodial symmetry) starts with an effective chiral
Lagrangian describing the Goldstone Bosons of the spontaneously broken \lr\
chiral symmetry. In the sense of the equivalence theorem \cite{kn:ET}, these
Goldstone Bosons represent the strongly-interacting longitudinal components of
the weak gauge bosons.

Instead, we wish to describe a theory in which an entire scalar-doublet $\phi$
is lighter than a TeV. We first consider theories in which the new dynamics is
{\it strongly} interacting and in which, typically, the Higgs is a composite
particle. Examples of such theories include the ``Composite Higgs'' models of
Georgi and Kaplan \cite{kn:gk}, as well as ``top-condensate'' and related
models \cite{kn:top}.

In these theories the usual ``power-counting'' rules of chiral perturbation
theory \cite{kn:gm,kn:gshort} apply: the coefficients of all
chirally-invariant interactions in the effective Lagrangian are expected to be
of order one if (1) there is an overall factor of $f^2 \Lambda^2$, (2) each
factor of the Higgs field appears with a factor of $1/f$ and (3) derivatives
appear with a factor of $1/\Lambda$.  Here $f$ is a measure of the amplitude
for producing the Higgs \cite{kn:gshort} and $\Lambda$ is the scale of
additional, heavy, strongly interacting particles.  For consistency, $\Lambda$
must be less than of order $4\pi f$ \cite{kn:gm}. As stated above, we are
interested in the situation that $\Lambda$ is of order one or a few TeV, but
we need not assume that $f$ is that large.

The effective Lagrangian describing the Higgs doublet in such a theory is
conveniently written in terms of the matrix $\Phi = (i\sigma_2 \phi^*, \phi)$,
which transforms\footnote{ As usual, $SU(2)_L$ will be identified with
  $SU(2)_{weak}$ and $SU(2)_R$ is the custodial $SU(2)$ symmetry whose
  $\tau_3$ component will be identified with hypercharge.} as $\Phi
\rightarrow L \Phi R^\dagger$ under \lr.  The most general custodially
symmetric Lagrangian to order momentum squared is
\be
\cL = {1 \over 2} \trace\, \left[ F \left( {\Phi^\dagger \Phi \over
    f^2} \right) \pdmu \Phi^\dagger \PDmu \Phi \right]
    - f^2 \Lambda^2 \trace\, G\left( {\Phi^\dagger \Phi \over f^2} \right),
\label{eq:newlag}
\ee
where $F$ and $G$ are arbitrary (dimensionless) functions analytic\footnote{In
  Composite Higgs models the non-derivative interactions of the Higgs doublet
  in the potential arise from small chiral-symmetry violating interactions and
  therefore, we must also require that coefficients in the expansion of $G$ be
  small compared to 1.} at $\Phi^\dagger \Phi=0$ and with $F(0)=1$.

In order for $\su$ breaking to occur, the potential $G$ must be minimized for
$\Phi \neq 0$. We may analyze the Lagrangian (\ref{eq:newlag}) by expressing
$\Phi$ in ``polar'' coordinates, $\Phi= \rho \yu/\sqrt{2}$.  Here $\rho$ is
real and positive, and $\yu$ is a special unitary matrix.  The Lagrangian to
order momentum squared may be written
\be
\cL = {1\over 2} \pdmu \rho \PDmu \rho + {\rho^2 \over 4}
A \left( {\rho^2 \over f^2}\right)
\trace\, (\pdmu \yu^{\dagger}\PDmu\yu) - \Lambda^2 f^2 B\left( {\rho^2 \over
  f^2}\right) ,
\label{eq:unshift}
\ee
where $A$ and $B$ are analytic functions related to the $F$ and $G$ above (in
particular, $F(0)=1$ implies $A(0)=1$). We have not included a function
multiplying the kinetic-energy term for the $\rho$ field because such a
function can always be eliminated by a redefinition of the $\rho$ field.
Writing $\rho = \langle \rho \rangle + H$, expanding around the true vacuum,
and keeping only the first few terms, we find
\be
\cL = \frac{1}{4} (v^{2} + 2 \xee v H + \xpr H^{2} + \xdpr
\frac{H^{3}}{6 v} )\; \trace\, (\pdmu \yu^{\dagger}\PDmu\yu) \;\;  + \;\;
\cL_{H}  \label{eq:Lagrangian}
\ee
where \xee\ , \xpr\, and \xdpr\ are unknown coefficients, $v^2 =
\vev^2 A(\vev^2/f^2) = (246 \ {\rm GeV})^2$, \yu\ contains the Goldstone Bosons
$w^{a}$\
\[
\yu = \exp \left(\frac{i \vw \cdot \pauli}{v} \right) \;\;\;\; , \;\;\;
       \trace \,  (\tau^{a} \tau^{b}) = 2 \delta^{a b}
\]
and $\cL_{H}$\ is the isoscalar Lagrangian
\be
\cL_{H} = \frac{1}{2} (\pdmu H)^{2} - V(H).
\ee
The leading terms in the scalar potential $V(H)$\ are
\be
V(H) = \frac{m^{2}}{2} H^{2} +
    \frac{\ctri v}{3 !}H^{3} + \frac{\ctes}{4 !}H^{4} \label{eq:pot}.
\ee

The power-counting rules imply that
\be
\xi,\xpr=1 + \cO\left({v^2\over f^2}\right)\; , \;
\xdpr = \cO\left({v^2\over f^2}\right)\label{eq:extra}
\ee
and
\be
\ctri, \ctes= {3m^{2} \over v^{2}}+\cO\left({v^2\over f^2}\right).
\label{eq:eextra}
\ee
We recover the usual linear sigma model in the limit that $f\rightarrow
\infty$.  For models in which the Higgs resonance is heavy (600 GeV or
higher), we expect $\Lambda$ to be of order one or a few TeV. In this case,
$v/f$ need not be small, and the $\xi$'s and $\lambda$'s can differ
substantially from their standard model values.

If, instead of being strongly coupled, the new dynamics is {\it weakly}
coupled, non-derivative and derivative interactions are treated in the same
way \cite{kn:gshort} and one writes an effective Lagrangian as an expansion in
$1/\Lambda$. The lowest order corrections to the standard Higgs Lagrangian
includes only dimension six operators
\be
   \sum_{i} \frac{c_{i}}{\Lambda^{2}} \cO_{i} \label{eq:lone}
\ee
where a minimal set of custodial $SU(2)_R$\  preserving $\{ \cO_{i} \}$\ is
\cite{kn:buchmuller}
\bear
\cO_{1} &=& \left(\phi^{\dagger}\phi\right)\,\pdmu\phi^{\dagger}\PDmu
  \phi \;-\; \frac{1}{4} \pdmu\left(\phi^{\dagger}\phi\right) \PDmu
\left( \phi^{\dagger}  \phi\right)          \nonumber\\
\cO_{2} &=& \left( \phi^{\dagger}\phi \right)^{3} \label{eq:d6o}
\eear
There are only two additional parameters here, instead of the five above. The
relationship between the $c_i$ and the parameters of eq.~(\ref{eq:Lagrangian})
is
\bear
\xee-1 &=& - \frac{3}{4}\; c_1 \; \frac{v^2}{\Lambda^2} \nonumber\\
\xpr-1 &=& 3\; c_1 \; \frac{v^2}{\Lambda^2} \nonumber\\
\xdpr &=& 12\; c_1 \; \frac{v^2}{\Lambda^2}  \nonumber\\
\ctri &=& \frac{3 m^2}{v^2} + \left[ \frac{3}{2} \left( \frac{m^2}{2
v^2}\right)
  c_1 - 6 c_2 \right] \frac{v^2}{\Lambda^2}  \nonumber\\
\ctes &=& \frac{3 m^2}{v^2} + \left[ 3 \left( \frac{m^2}{2 v^2}\right)
  c_1 - 36 c_2 \right] \frac{v^2}{\Lambda^2}  \label{eq:new1}
\eear

We may now consider the phenomenology of a non-standard Higgs resonance. At
tree level, the Higgs boson decay width to Goldstone Bosons is
\be
\Gamma_{H}^{(0)}=\frac{3m^{3}}{32\pi v^{2}}\xi^{2} \label{eq:treewidth}
\ee
Note that \xee\ is the only parameter which appears \cite{kn:kosower,kn:tasi}.

The other parameters in eq.~(\ref{eq:Lagrangian}) appear at one-loop. As
usual, loops induce infinities which can be absorbed in the effective
Lagrangian
in the traditional way \cite{kn:gm,kn:gl}: namely the infinities associated
with non-derivative interactions are absorbed in the renormalization of the
scalar self couplings in eq.~(\ref{eq:pot}), while the ones associated with
vertices involving derivatives are absorbed in the counterterms of order
$p^{4}$. In general, these introduce further unknown parameters in our
amplitudes. We compute the leading corrections in the $\overline{MS}$
scheme, setting the $\cO(p^4)$ counterterms to zero when the renormalization
scale $\mu$ is equal to $\Lambda$.  These results include the so-called
``chiral logarithms'', which are the leading contributions if $p^2/\Lambda^2$
is sufficiently small \cite{kn:pagels}, and in any case are expected to be
comparable to the full $\cO(p^{4})$\ corrections \cite{kn:gm}. In addition,
when the parameters take the values of the linear sigma model ($f \rightarrow
\infty$ in eqs. (\ref{eq:extra}) and (\ref{eq:eextra})), the $\mu$-dependence
disappears (as it must for a renormalizable theory) and we show that our
results reduce to those previously computed in the standard Higgs model
\cite{kn:marciano,kn:dawson}.

The one-loop corrections to the Higgs boson decay width in
eq.~(\ref{eq:treewidth}), written as $\Gamma_{H}= \Gamma_{H}^{(0)}+
\Gamma_{H}^{(1)}$, are
\bear
\lefteqn{\frac{\Gamma_{H}^{(1)}}{\Gamma_{H}^{(0)}} = \frac{1}{8\pi^{2}}
  \left\{ \frac{m^{2}}{v^{2}}(1+L)+\frac{\xpr\ctri}{2\xi}\left[
  \frac{\pi}{\sqrt{3}}-1 \right] + \frac{\ctri^{2}v^{2}}{4m^{2}}
   \left[ 1-\frac{2\pi\sqrt{3}}{9}\right]  +\frac{\xdpr
      m^{2}}{\xi v^{2}} L \right. } \nonumber\\ & &
\left. \mbox{}+ \xpr \frac{3m^{2}}{2 v^{2}} \left[\frac{1}{3} +L\right]
      +  \frac{\xi^{2}m^{2}}{2v^{2}} \left[\frac{\pi^{2}}{6}
  -4-L \right]- \frac{\xi\ctri}{2}\left[\pi\sqrt{3}-3-\frac{2\pi^{2}}{9}
  \right]   \right\} \label{eq:hwidth}
\eear
where $L=1-\ln (m^{2}/\mu^{2})$. While this result is $\mu$-dependent, we can
estimate the effect of higher-order interactions by setting $\mu=\Lambda$. In
the linear sigma model limit our calculation reproduces the one-loop result of
ref.~\cite{kn:marciano} where
\be
\frac{\Gamma_{H}^{(1)}}{\Gamma_{H}^{(0)}} = \frac{m^{2}}{2\pi^{2}
v^{2}} \left( \frac{19}{16} -\frac{3\sqrt{3} \pi}{8} + \frac{5
\pi^{2}}{48} \right)
\ee
To get a feeling for the magnitude of the deviations, we set the parameters of
eq. (\ref{eq:hwidth}) to the values\footnote{These values correspond to a
  representative choice of parameters in the model described in
  \cite{kn:vklong}.}
\be
m=750\,\gev\;\; ,\; \xi=0.6\;\; ,\; \xpr=-0.26\;\; ,\;\xdpr=0.66\;\; ,\;
\ctes= 3.34\;\; ,\; \ctri= 20 \;\; ,\; \Lambda=2\, \mbox{TeV}
\label{eq:values}
\ee
The one-loop contribution increases the width by $24 \%$\ while the
corresponding increase for the linear sigma model is only $8 \%$.

We now investigate the behavior of such a non-standard Higgs resonance in
longitudinal Goldstone Boson scattering. By expanding the Lagrangian in
eq.~(\ref{eq:Lagrangian}), the tree-level amplitude for \wz\ is
\be
\cA_{tr} = \frac{s}{v^{2}} - \left(\frac{\xi^{2}}{v^{2}}\right)
 \frac{s^{2}}{s-m^{2}- \Sigma(s)}  \label{eq:tree}
\ee
In order to consistently count powers of $\lambda$ ($\equiv m^2/2v^2$) in the
resonance region, we use $\Sigma(s)= i {\rm Im} \Pi_H^{one-loop}(s)$, where
$\Pi_H$ is the Higgs boson self-energy function.  The first term comes from
the gauge boson contact interaction and corresponds precisely to the low
energy theorems \cite{kn:chano}, while at somewhat higher energies deviations
from the standard Higgs model emerge.

Next, we compute the one-loop contributions to the \wz\
scattering amplitude. The full analytical expression will be presented in
ref. \cite{kn:vklong}.  Here we provide the result for the $I=J=0$\ partial
wave
defined as
\be
a_{00}=\frac{1}{16\pi s} \int_{-s}^{0} dt\, \left[ 3\cA(s,t,u) + \cA(t,s,u)
+ \cA(u,t,s)\right]
\ee
where $\cA(s,t,u)$\ is the \wz\ amplitude, which includes the tree-level
expression given in eq.~(\ref{eq:tree}) (here with $\Sigma(s) =
\Pi_H^{one-loop}(s) + i {\rm Im} \Pi_H^{two-loop}(s)$) and the one-loop
corrections calculated in the $\overline{MS}$\ scheme.

At energies small compared to the mass of the Higgs, the one-loop amplitude
is :
\be
\cA(s,t,u) = \frac{s}{v^{2}} + \frac{1}{(4\pi v^{2})^{2}}\; \cM
\ee
\bear
\cM & = & \frac{s^{2}}{2} \ln\frac{\mu^{2}}{-s} + \frac{1}{6} t(s +2t)
     \ln \frac{\mu^{2}}{-t} + \frac{1}{6} u(s+ 2u)\ln\frac{\mu^{2}}{-u}
\nonumber\\*[0.1 cm]  & & \mbox{} +
   s^{2}\, P \; + \; Q\, (t^{2}+u^{2}) \; + \, R\ln \frac{m^{2}}{\mu^{2}}
\eear
where
\bear
 P & = & \frac{5}{9} - \frac{49}{18} \xi^{4} +
\frac{\xi^{3}\ctri}{4\lambda} + \frac{\xi^{2}}{4} \left(
  \frac{\ctes}{\lambda} -  \frac{\lambda_{3}^{2}}{2 \lambda^{2}} \right)
  + \xi^{2}  \left( \frac{5}{2} \xpr -\frac{37}{18} \right)
\nonumber\\*[0.1 cm]  & &  \mbox{} +
  \xi \left( \frac{\xpr\ctri}{2\lambda} + 2\xdpr -\frac{\ctri}{2\lambda}
 \right)  + \xpr -
  \frac{\ctes}{4 \lambda} + \frac{\lambda_{3}^{2}}{8 \lambda^{2}} \left(
   \frac{\pi}{\sqrt{3}} -1 \right) \\*[0.1 cm]
Q  & = &  \frac{13}{18} -  \frac{11}{9}\xi^{2} + \frac{5}{18} \xi^{4} \\
R  & = & s^{2}\, \left[ \frac{\ctes}{4 \lambda} - \xpr\left( 1+
  \frac{\xpr}{2} \right) + \xi \left( \frac{\ctri}{2 \lambda} -2 \xdpr
\right) + \xi^{2} \left(
 \frac{7}{6} - \xpr - \frac{\ctes}{4 \lambda}\right) \right.
\nonumber\\*[0.1 cm] & & \mbox{}  - \left.
    \frac{\xi^{3} \ctri}{2  \lambda}  + \frac{5}{3} \xi^4 \right] \; +
  \; \frac{\xi^2}{3} \left( 2 - \xi^2 \right) \left(t^{2} +u^2 \right)
\eear
This reduces to the appropriate expression in the linear-sigma model
limit \cite{kn:dawson}.

In Fig.~1 we show the modulus of the $I=J=0$\ partial wave as a function of
$\sqrt{s}$ for our benchmark parameter values eq.~(\ref{eq:values}). The peak
in the one-loop curve is $\sim 9 \%$ lower (as well as being wider) than the
tree-level peak. The corresponding curves for the SM Higgs are shown in
Fig.~2.  Note that both the tree-level and one-loop amplitudes in the linear
and nonlinear models violate (or, at least, do not saturate) unitarity on the
peak. This is an indication that perturbation theory is not very accurate on
the peak: even though the amplitudes obey unitarity to the appropriate order
in perturbation theory \footnote{This has been studied in detail in the linear
  sigma model in \cite{kn:willval}.}, the higher order corrections are large.

Qualitatively, however, for gauge-boson scattering below a TeV, the width
appears to be the most important feature differentiating a standard from a
non-standard Higgs resonance\footnote {For SSC phenomenology, the potentially
  non-standard coupling of the Higgs boson to the top-quark will also be
  important since it will affect the Higgs production rate.}.

The sharp fall in the amplitude in the region above the peak in Fig.~1 can be
understood by noticing that for $\xi < 1$\ the tree amplitude in
eq.~(\ref{eq:tree}) vanishes at some energy greater than $m^2$ (if one does
not include a finite width). This only signals that higher order effects are
expected to be significant there. Also, far above the peak the amplitude
presented is not trustworthy due to the breakdown of the expansion in
powers of $1/\Lambda$.

So far, we have ignored all gauge corrections. These are small in longitudinal
gauge boson scattering, as shown by the Equivalence Theorem. However, our
non-standard Higgs boson does contribute to electroweak radiative corrections,
including the Peskin-Takeuchi parameter S \cite{kn:peskin}
\be
S= - 16\pi \,\frac{d}{dq^{2}} \, \left. \Pi_{3Y}(q^{2})\right|_{q^{2}=0}
\ee
Again, we will only compute the leading corrections
\cite{kn:golden} to $S$.

Since $\Sigma$\ transforms under \lr\ as $\Sigma\rightarrow L\Sigma
R^{\dagger}$\ the gauge bosons are included in the theory by replacing
the ordinary derivative by the covariant one
\be
D_{\mu}\Sigma = \partial_{\mu}\Sigma + i\frac{g}{2}\vec{\tau}\cdot
\vec{W_{\mu}}\Sigma - i\frac{g^{\prime}}{2} B_{\mu} \Sigma \tau_3
\ee
The diagram shown in Fig.~3 contributes to $\Pi_{3Y}$. We calculate this
diagram using dimensional regularization ($\overline{MS}$) and absorb the
infinities in the $p^{4}$\ counterterms.  Then, after subtracting the
contribution from ``known'' physics (the Standard Model with a Higgs boson of
mass $m_0$, usually taken to be 1 TeV \cite{kn:peskin}) we obtain
\be
S=\frac{1}{12\pi} \left\{
\ln \frac{m^2}{m^2_0} +\left( 1-\xi^{2} \right) \left[\frac{1}{6} -\ln
\frac{m^{2}}{\mu^{2}} \right] \right\}
\ee
where $\mu$ is the renormalization scale.  For our benchmark parameters
(\ref{eq:values}), $m_0=$ 1 TeV, and $\mu = \Lambda$ we find S = +0.021.  If
$\xi$ were greater than $1$, the contribution to S could even be
negative, and slightly larger in absolute value than the corresponding
contribution in the standard model with the same Higgs mass.

\bigskip

In this letter we have constructed an effective Lagrangian description
of a non-standard Higgs boson. At lowest order in momentum and up to one-loop
this description includes 5 new parameters. We have calculated the leading
corrections to the Higgs boson width, \wz\ scattering, and the
Peskin-Takeuchi $S$ parameter.

We have found that the most prominent effects are due to one new parameter,
namely \xee , which is directly related to the Higgs boson width.  The
one-loop corrections do not qualitatively change the features of longitudinal
gauge boson scattering, although a detailed analysis can only be done in a
specific model.  If a ``Higgs'' is discovered, it will be important to see
whether or not its properties are those predicted by the standard model.  It
remains to be seen how well the parameter \xee\ can be measured.

\section*{Acknowledgments}

We would like to thank E.~H.~Simmons, H.~Georgi, M.~Golden, D.~Kominis, and
S.~Selipsky for useful discussions and suggestions.  R.S.C.  acknowledges the
support of an Alfred P. Sloan Foundation Fellowship, an NSF Presidential Young
Investigator Award, a DOE Outstanding Junior Investigator Award, and a
Superconducting Super Collider National Fellowship from the Texas National
Research Laboratory Commission.  This work was supported in part under NSF
contract PHY-9057173 and DOE contract DE-FG02-91ER40676, and by funds from the
Texas National Research Laboratory Commission under grant RGFY92B6.

\newpage
\noindent
{\Large\bf Figure captions}
\vspace{0.5cm}

\begin{itemize}
\item
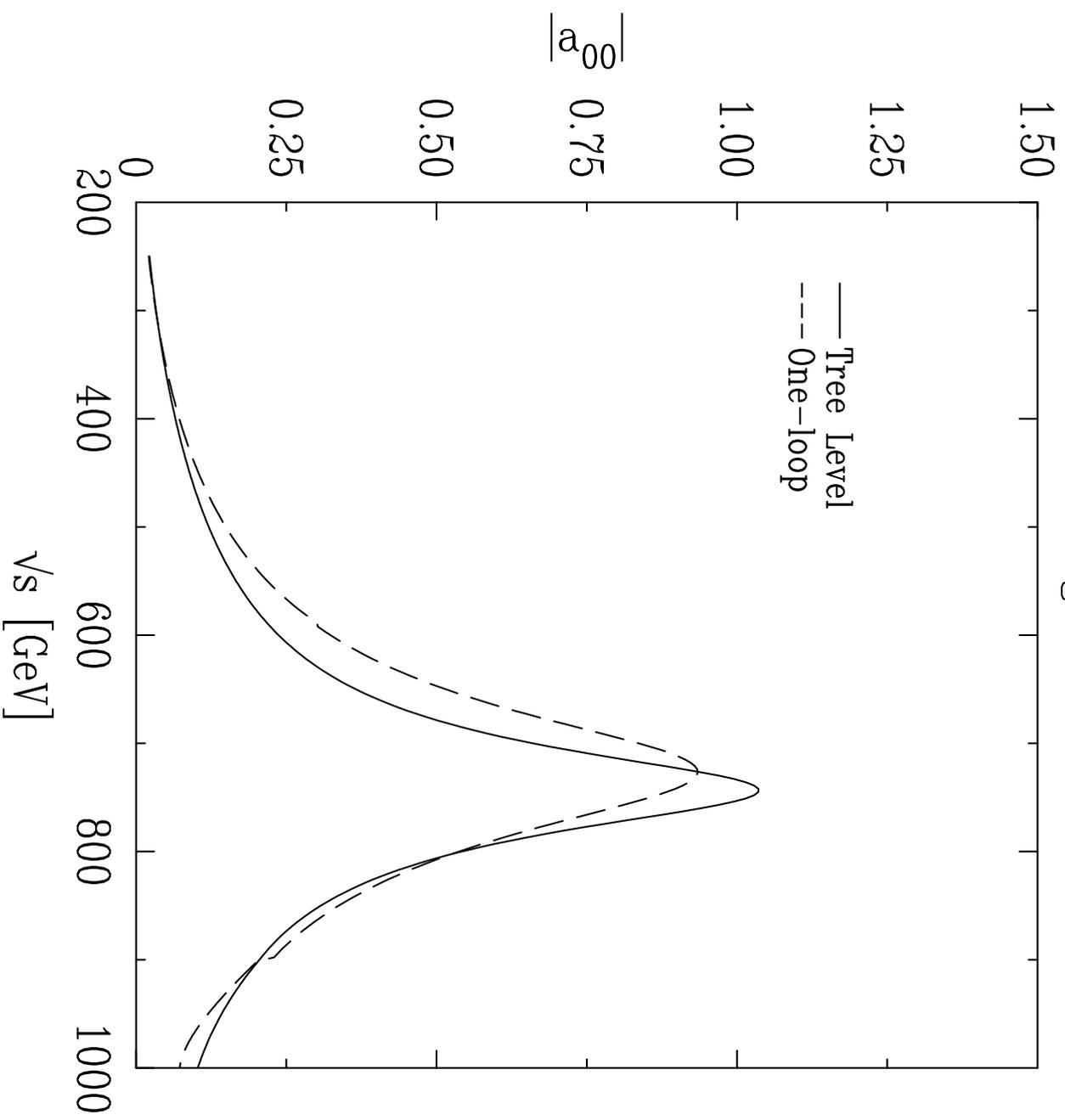
  {\it Figure 1} : The absolute value of the $I=J=0$\ partial wave for gauge
  boson scattering with a non-standard Higgs boson (using the values shown in
  eqn. (\ref{eq:values})) at tree-level (solid) and one-loop (dotted).
\item
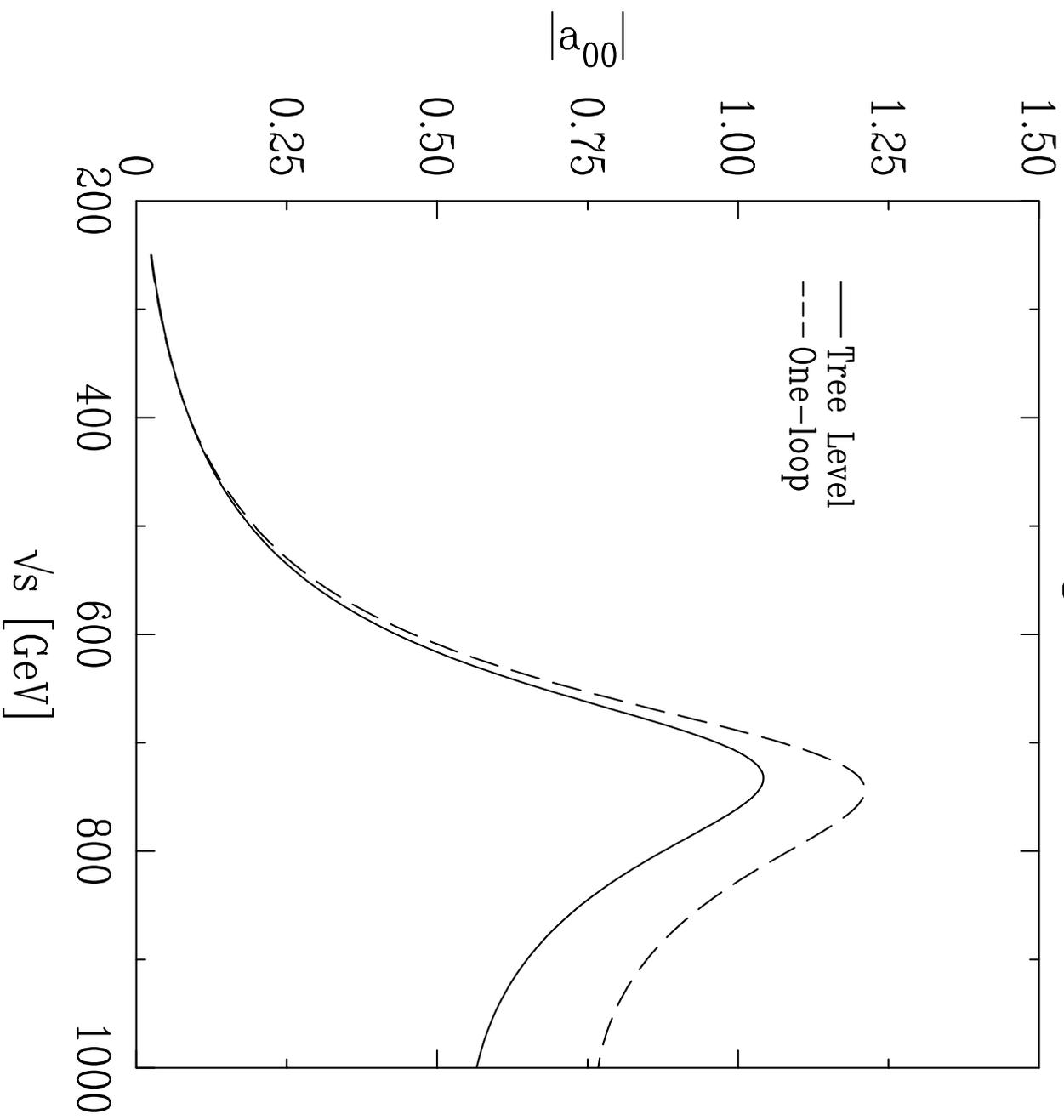
  {\it Figure 2} : The absolute value of the $I=J=0$\ partial wave for gauge
  boson scattering with a standard Higgs boson at tree-level (solid) and
  one-loop (dotted).
\item
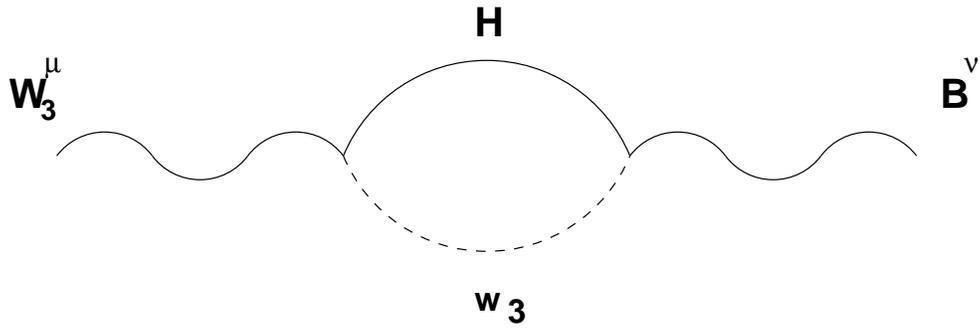
  {\it Figure 3} : The one-loop diagram contribution to the Peskin-Takeuchi
  {\bf S} parameter.
\end{itemize}

\end{document}